\documentclass[11pt]{article}
\usepackage{hyperref}
\begin{document}
\title{Rodless Weissenberg Effect }
\author{Enrique Soto*, Oscar R. Enr\'iquez, Roberto Zenit and Octavio Manero \\
\\\vspace{6pt} *The W.G. Pritchard Laboratories, Department of Mathematics\\Penn State University, University Park, PA 16802, USA
\\ \\ Instituto de Investigaciones en Materiales \\ Universidad Nacional Aut\'onoma de
M\'exico\\ Apdo. Postal 70-360, M\'exico D.F. 04510, M\'exico }
\maketitle
\begin{abstract}

The climbing effect of a viscoelastic fluid when stirred by a
spinning rod is well documented and known as Weissenberg
effect(Wei et al, 2006). This phenomenon is related to the
elasticity of the fluid. We have observed that this effect can
appear when the fluid is stirred without a rod. In this work, a
comparison of the flow around a spinning disk for a Newtonian and
a non-Newtonian liquids is presented
(\href{http://hdl.handle.net/1813/11467}{Video}).  The flow is
visualized with ink and small bubbles as fluid path tracers. For a
Newtonian fluid, at the center of the spinning disk, the fluid
velocity is directed towards the disk (sink flow); on the other
hand, for a viscoelatic liquid, a source flow is observed since
the fluid  emerges from the disk. The toroidal vortices that
appear on top of the disk rotate in opposite directions for the
Newtonian and non-Newtonian cases. Similar observations have been
reported for the classical rod climbing flow (Siginer, 1984 and
Escudier, 1984). Some authors have suggested that this flow
configuration can be used to determine the elastic properties of
the liquid (Escuider, 1984 and Joshep, 1973).

\end{abstract}
\begin{enumerate}
\item Wei, J., Li, F., Yu, B. and Kawaguchi, Y. \emph{Swirling
Flow of a Viscoelastic Fluid Whit Free Surface-Part I:
Experimental Analysis of Vortex Motion by PIV}, Journal of Fluids
Engineering, \textbf{128}, 69-76 (2006).

\item Siginer, A. \emph{General Weissenberg Effect in Free Surface
Rheometry Part I: Analytical Considerations}, Journal of Applied
Mathematics and Physics (ZAMP), \textbf{35}, 545-548 (1984).

\item Escudier, M.P. \emph{Observations of the Flow Produced in a
Cylindrical Container by a Rotating Endwall}, Experiments in
Fluids, \textbf{2}, 189-196 (1984).

\item Joseph, D.D., Beavers, G.S. and Fosdick, R.L. \emph{The Free
Surface on a Liquid Between Cylinders Rotating at Different
Speeds, Part II}, Arch. Rat. Mech. Anal. \textbf{49}, 381-401
(1973).

\end{enumerate}

%
\end{document}